\documentclass{article}
\usepackage{ijcai19}

\usepackage{fancyhdr}
\usepackage{times}
\usepackage{soul}
\usepackage{url}
\usepackage[hidelinks]{hyperref}
\usepackage[utf8]{inputenc}
\usepackage[small]{caption}
\usepackage{graphicx}
\usepackage{epsfig}
\usepackage{multirow}
\usepackage{amsmath}
\usepackage{array,booktabs}
\usepackage{algorithm}
\usepackage{algorithmic}
\usepackage{enumitem} %for list indentations
\usepackage{multirow}

\newcommand {\otoprule}{\midrule [\heavyrulewidth]} 
\newcolumntype {+}{ >{\global\let\currentrowstyle\relax}}
\newcolumntype {^}{ >{\currentrowstyle }}
 \newcommand {\rowstyle}[1]{\gdef\currentrowstyle{#1} %
 #1\ignorespaces
 }
\newcommand{\tabhead}{\rowstyle{\bfseries}}

\usepackage{xcolor}
\definecolor{colchristin}{rgb}{0.0, 0.0, 1}

\definecolor{colandrea}{rgb}{0.0, 0.5, 0.5}

\newcommand{\HH}{$C_{0.97}^{high}$}
\newcommand{\HL}{$C_{0.97}^{low}$}
\newcommand{\HN}{$C_{0.97}^{no}$}
\newcommand{\MH}{$C_{0.76}^{high}$}
\newcommand{\ML}{$C_{0.76}^{low}$}
\newcommand{\MN}{$C_{0.76}^{no}$}
\newcommand{\LH}{$C_{0.03}^{high}$}
\newcommand{\LL}{$C_{0.03}^{low}$}
\newcommand{\LN}{$C_{0.03}^{no}$}

\newcommand{\High}{$C_{0.97}$}
\newcommand{\Medium}{$C_{0.76}$}
\newcommand{\Low}{$C_{0.03}$}

\title{How model accuracy and explanation fidelity influence user trust in AI}

\author{
Andrea Papenmeier$^{1,2}$
\and
Gwenn Englebienne$^2$ \and
Christin Seifert$^2$
\affiliations
$^1$GESIS - Leibniz Institute of the Social Sciences, Germany\\
$^2$University of Twente, Netherlands\\
\emails
andrea.papenmeier@gesis.org,
g.englebienne@utwente.nl,
c.seifert@utwente.nl
}

\begin{document}

\maketitle
\thispagestyle{fancy}

\begin{abstract}
  	Machine learning systems have become popular in fields such as marketing, financing, or data mining. While they are highly accurate, complex machine learning systems pose challenges for engineers and users. Their inherent complexity makes it impossible to easily judge their fairness and the correctness of statistically learned relations between variables and classes. Explainable AI aims to solve this challenge by modelling explanations alongside with the classifiers, potentially improving user trust and acceptance. However, users should not be fooled by persuasive, yet untruthful explanations. We therefore conduct a user study in which we investigate the effects of model accuracy and explanation fidelity, i.e. how truthfully the explanation represents the underlying model, on user trust. Our findings show that accuracy is more important for user trust than explainability. Adding an explanation for a classification result can potentially harm trust, e.g. when adding nonsensical explanations. We also found that users cannot be tricked by high-fidelity explanations into having trust for a bad classifier. Furthermore, we found a mismatch between observed (implicit) and self-reported (explicit) trust.
\end{abstract}

%--------------------------------------------------------------------------------------
\section{Introduction}
The need for explanations of machine learning algorithms has been identified in the past \cite{richardson2018survey,goodman16eu} and led to the emergence of the research field of explainable artificial intelligence (xAI). Several researchers argue that explanations have a positive effect on user trust \cite{biran2017explanation,glass2008toward,preece2018asking,vorm2018assessing} and that inappropriate trust impairs the human machine interaction \cite{preece2018asking,ribeiro2016should}. However, explanations do not necessarily have to deliver accurate information about the machine learning algorithm. Yet, untruthful explanations with low fidelity to the machine learning model can appear plausible to the user \cite{lipton2016mythos}. It has not yet been established how characteristics such as fidelity of an explanation impact user trust.\newline
We therefore investigate how  varying explanation fidelity influences the user's trust into an automatic decision system. Using the scenario of a ``social media administrator" with the task to detect offensive language in Tweets, we develop three machine learning classifiers able to process textual input and classify the texts into ``offensive" and ``not offensive" classes at varying levels of accuracy. Furthermore, we implement and validate the automatic generation of explanations at high fidelity and low fidelity levels. We measure the trust and perceived understanding in a user study with 327 participants in order to compare different classifier-explanation combinations. Our research was driven by the following questions:
\begin{enumerate}[label=RQ\arabic*:,itemindent=6pt,noitemsep,topsep=0pt,parsep=0pt,partopsep=0pt]
	\item What influence does the accuracy of an automatic decision system have on user trust?
	\item How do the presence and the level of fidelity of explanations influence user trust?
\end{enumerate}
%\vspace{0.1cm}\newline
%\textit{RQ 1: What influence does the accuracy of an automatic decision system have on user trust?}\newline
%\textit{RQ 2: How do the presence and the level of fidelity of explanations influence user trust?}\vspace{0.1cm}\newline
%\christin{the following is too detailed for the introduction. Might be quite confusing if one does not know what "our minimal explanations are". Suggestion: move this to the back. but leave the general statements here "our key findings .. influence .. variety of ways" and ". "In general .. most decisive for .." And add that the picture of the fidelity of explanaitons is more complex.}
Our key findings show that explanations affect user trust in a variety of ways, depending on the overall accuracy of the system, the fidelity level of the explanation, and the user's level of consciousness. In general, the systems' accuracy levels were most decisive for user trust: the higher the accuracy, the higher the user's trust. The influence of explanation fidelity differs depending on the model accuracy: We see that for systems with medium accuracy, a high-fidelity explanation does not harm user trust, while a low-fidelity explanation does. Yet, for a system with high accuracy, any explanation (high-fidelity as well as low-fidelity) leads to a decrease in trust. We conclude that the interplay between explanation fidelity and user trust is more complex than pictured in literature to date. Furthermore, our findings show a discrepancy between how users act and what users report, which should be taken into account when evaluating user trust.\newline
With our research, we contribute empirical evidence of the relation between accuracy, fidelity, and user trust to the xAI community. Other than related research, we focus on the practical implications of explainability and their effect on the relationship with the user. Furthermore, we test an observational measure of trust as an objective method complementing traditional self-reported trust questionnaires.\newline
In this paper, we first review the existing literature on explanations and user trust in AI. We then derive the structure for a user study (section \ref{sec:stud_design} and \ref{sec:experiment}) aiming to test the influence of explanation fidelity and classifier accuracy on user trust. Finally, we present and discuss the results in section \ref{sec:res_disc}.

%--------------------------------------------------------------------------------------
\section{Related Work}
\label{sec:rel_work}
Artificial intelligence and machine learning algorithms are nowadays employed in a variety of areas. In safety-critical applications such as terrorism detection \cite{ribeiro2016should} or autonomous robotics \cite{richardson2018survey}, faulty behaviour needs to be avoided at all costs. Furthermore, machine learning systems treating sensitive data such as credit ratings \cite{domingos2012few} or health applications \cite{goodman16eu} need to communicate what brought about a single decision. \cite{goodman16eu,wachter2017right,selbst2017meaningful} discuss a ``right to explanation" or ``right to information" as a consequence of the General Data Protection Regulation (GDPR) introduced in the EU in 2018. Overall, those systems not only need to be right in a high number of cases, but right for the right reasons \cite{preece2018asking}.

\subsection{Explanations in AI}
When being confronted with new information, humans incorporate them in mental models. Explanations are a tool to build and refine inner knowledge models \cite{miller2018explanation}. For an engineer working on a machine learning system, understanding underlying principles and consequences of the system's behaviour is a necessary step in designing a system that is ``right for the right reasons" \cite{preece2018asking}. On the user side, explanations have a positive effect for the ability to predict the system's performance correctly \cite{biran2017explanation}. \cite{ribeiro2018anchors} found that explanations increase the user's ability to predict the classifier decision, while decreasing the time needed to reach a judgement. Their within-subject study design, however, could have led to familiarisation and hence an overrating of explanations.\newline
In recent years, machine learning algorithms show a trend towards increasing accuracy, but also increasing complexity. In general, the higher the accuracy and complexity, the lower the explainability \cite{chen2018learning,richardson2018survey}. An interpretable machine learning system is either inherently interpretable (e.g. decision trees, linear models \cite{biran2017explanation}), or is capable of generating descriptions understandable to humans \cite{lipton2016mythos}. \cite{lipton2016mythos} points out that a retrospectively added explanation does not guarantee fidelity, ``however plausible they appear".\newline
To achieve explainability, \cite{chen2018learning} developed an add-on explanation system for texts based on mutual information analysis and measure the explanations fidelity to the underlying model with good results. \cite{feng2018pathologies} went a step further with an image classification system and add-on textual explanatory mechanism. However, they also show that their high-fidelity explanations are nonsensical for humans. In human-human explanations, people tend to question underlying principles of events by comparing it to known concepts. ``Why A, why not B?" is a common question during this thought process \cite{miller2018explanation}. \cite{chen2018learning} suggests showing reference cases in automatic decision systems: similar cases with a different predicted class, or dissimilar cases (counterfactuals) \cite{hendricks2018generating}. Approximating elements of an opaque system is another method of achieving interpretability. \cite{domingos2012few} argues that most high-dimensional real-world application data is ``concentrated on or near a lower-dimensional manifold" and suggests dimension reduction techniques to reduce the complexity of a system to a human-comprehensible level. \cite{chen2018learning} suggests salience map masks on input features to point the attention towards features that are decisive in a sample, e.g. single words in texts. \cite{goodman16eu} suggests a ``minimum explanation", showing at least how input features relate to the prediction of a classifier.\newline
%\christin{one - two sentences how this all relates to your work.}
In summary, \cite{chen2018learning}'s model-agnostic explanations combined with \cite{goodman16eu}'s definition of minimum explanations provide a basis for examining the influence of explanation fidelity and model accuracy. However, following \cite{lipton2016mythos}, explanation fidelity needs to be validated computationally.

\subsection{Trust in AI}
Literature suggests that insights into the system functioning and decision process increase trust \cite{biran2017explanation,glass2008toward,preece2018asking,vorm2018assessing}. In the field of computer science, most definitions agree in that trust relates to the assurance that a system performs as expected \cite{mohammadi2013trustworthiness}. Since trust is placed in an agent by another agent, it is not an objective measure but a subjective experience of an individual \cite{mohammadi2013trustworthiness}. \cite{korber2018theoretical} developed a trust metric for automated systems based on a model of human-human trust. It consists of 19 self-report items measuring the trust factors reliability, predictability, the user's propensity to trust, as well as the attitude towards the system's engineers and the user's familiarity with automated systems.\newline
For trust in automatic classification systems, misclassifications (i.e. the system's prediction does not correspond to the user's prediction) play a special role, as they can lead to a decrease in user trust \cite{glass2008toward}. \cite{vorm2018assessing} reports ``willingness to accept a computer-generated recommendation" as an observable sign for trust. \cite{yu2017user} found that users are able to detect the accuracy of a classifier without being told explicitly, and adjust their trust accordingly. \cite{cramer2008effects} tested the effects of transparency on user perception, finding a correlation between perceived understanding and trust, but no evidence for a direct influence of transparency on trust. They hypothesise that transparency also discloses system boundaries and unfulfilled preferences, ultimately cancelling out any positive effects. \cite{langer1978mindlessness} found in a user study that the pure presence of an explanation, regardless of the content, can make a difference in how people react to requests. Without explanation, humans complied significantly less with a request than in cases where an explanation was given. They compared nonsensical and meaningful explanations, but found only little difference in their power of persuasion \cite{langer1978mindlessness}. They explain this behaviour with the state of ``mindlessness", triggering an automatic script ``comply if reason is given", no matter the given reason. The mindless state, however, is revoked if complying leads to stronger consequences. In an attentive state, the explanation does make a difference: People were more likely to comply when an informative explanation was given, as compared to a nonsensical one \cite{langer1978mindlessness}.\newline
%\christin{one - two sentences how this all relates to your work.}
Overall, evaluating trust implies measuring the subjective experience of users. Since \cite{langer1978mindlessness} observed a difference in user trusting behaviour between a ``mindful" and a ``mindless" state, trust should be evaluated both subjectively and objectively, e.g. using a questionnaire and observation.

%--------------------------------------------------------------------------------------
\section{Study Design}
\label{sec:stud_design}
As trust is a subjective experience, it must be evaluated in a user study. We use the following scenario for a user study: the social media presence of a company that targets teenagers and young adults (15-20 years old). The use case task is to identify offensive texts with the support of a machine learning system. To measure the influence of accuracy and explanation fidelity on user trust, we establish 9 conditions: three classifiers (high, medium, low accuracy), each with three explanation types (high-fidelity, low-fidelity, no explanation), see table \ref{tab:conditions}. To avoid learning and familiarisation effects, we use a between subject design, with each participant being assigned to one condition at the beginning of the survey. As trust builds during repeated interaction \cite{rempel1985trust}, we show participants a subset of 15 Tweets. We construct 10 disjunct subsets (cf. sec~\ref{sec:experiment}), to reduce the impact of specific wording or topics. At the start of the survey, each participant is randomly assigned to one subset.
\begin{table}[h]
	\centering
	\setlength\extrarowheight{2pt}
	\begin{tabular}{+l^l^l^l^l}
	\toprule
		&& \multicolumn{3}{c}{\textbf{Classifier Accuracy}} \\
  && high & medium & low \\ \otoprule
 \multirow{3}{*}{\rotatebox[origin=c]{90}{\textbf{Explan.}}}		
 	& high-fidelity & \HH & \MH & \LH \vspace{2pt}\\
	& low-fidelity & \HL & \ML & \LL \vspace{2pt}\\
	& no & \HN & \MN & \LN\\ \bottomrule
	\end{tabular}
	\caption{Classifier-explanation conditions}
	\label{tab:conditions}
\end{table} 

\subsubsection*{Apparatus \& Procedure}
The user study is set up as an online study on the soSci platform\footnote{https://www.soscisurvey.de/en/index accessed on 24.02.2019}. Participants are asked to access the survey via an online link on their private device, with small screens (e.g. smartphones) being excluded to ensure proper image scaling. Consistent with the use case scenario, screenshots of a fictive social media management platform show the input texts, decisions and explanations. The screenshots have a ratio of 900px (width) to 253px (height).\newline
The study consists of three blocks. In the first block, the participant is asked to manually classify 15 Tweets as offensive or not offensive. The second block introduces the automatic decision system, asking to classify 15 ``very similar" Tweets, which are in fact identical to those in the first block. The Tweets are pre-classified and displayed according to one of the 9 conditions. Finally, the last block contains questions to measure perceived understanding, trust (including an attention check), and the demographic background. The survey was tested in a pilot with 11 participants.

\subsubsection*{Measures \& Analysis}
We measure perceived understanding and trust quantitatively. For perceived understanding, we ask three statements to be rated on a 5-points Likert scale and take the average as a single score per participant. To measure trust, we observe how the system influences the participant's judgement by comparing the manual classifications of the first (without system) and second block (with system). We define changing a classification in favour of the system's prediction but away from the truth as a sign for being convinced and trusting the system. The opposite behaviour (changing towards the truth but away from the system's prediction) is interpreted as a sign for mistrust. 
We normalise the number of changes by the number  possibilities to see the behaviour in question 
(e.g. a highly accurate classifier offers only once the possibility to contradict the truth in favour of its prediction). As a subjective, self-reported trust measure, we use the questionnaire of \cite{korber2018theoretical}, taking the mean score over all 19 items for a single trust score per participant. We use the two-sided Mann-Whitney U test with Bonferroni correction to compare two score samples.

\subsubsection*{Participants}
Participants were recruited via the science crowdsourcing platforms Prolific\footnote{https://prolific.ac accessed on 24.02.2019} and SurveyCircle\footnote{https://www.surveycircle.com accessed on 24.02.2019}. In total, 327 participants took part in the main user study with an average age of 29.4 years (SD=8.8), with 56\% females and 43\% males. Two participants reporting the third gender. 57\% self-assessed their English as equivalent to a native speaker, but all participants claimed to be fluent in English. 41 data points were invalidated due to failed attention check and survey completion level, resulting in 286 valid cases.

%--------------------------------------------------------------------------------------
\section{Experiment}
\label{sec:experiment}
\subsubsection*{Dataset}
We use a dataset of offensive language and hatespeech\footnote{https://github.com/t-davidson/hate-speech-and-offensive-language accessed on 24.02.2019} provided by \cite{davidson2017automated}. It contains Tweets labelled by at least 3 annotators, of which we use only those data points with an inter-annotator agreement of 100\%. The final dataset contains 4324 Tweets with a class balance of 1:1. We randomly split the data set into training (80\%) test (20\%). The Tweets are preprocessed with a conversion to lower cases, common contraction solving (e.g. ``we're"), deletion of retrospectively added signifiers (e.g. ``RT" indicating a Re-Tweet), deletion of non-alphabetic characters (all besides hashtags), and replacement of URLs and user names by dummy handles. The texts are tokenized on whitespaces. 
%The following Tweet:\newline
%``@WBUR: A smuggler explains how he helped fighters along the ``Jihadi Highway": http://t.co/UX4anxeAwd"\newline
%is therefore processed into:\newline
%@username a smuggler explains how he helped fighters along the jihadi highway http://website.com/website\newline

\subsubsection*{Classifiers}
For the system with \textbf{high accuracy}, we adopt the setup used by \cite{chen2018learning}. They use a convolutional neural network (CNN) for sentiment analysis. We implement the CNN using the \textit{Keras}\footnote{https://keras.io accessed on 24.02.2019} Python library. \High achieves an accuracy of 0.97 on the test set. 
For the classifier with \textbf{medium accuracy} (\Medium) we adapt the approach of~\cite{davidson2017automated}, which uses logistic regression to identify offensive language and hate speech, achieving an F1-score of 0.9 on their test set. The logistic regression classifier is implemented with the \textit{scikit-learn}\footnote{https://scikit-learn.org accessed on 24.02.2019} Python library, with an L-BFGS optimiser. We adjust all positive coefficients of \Medium to a value of 1.0 and all negative to -1.0 to reach the final (medium) accuracy of 0.76 on the test set. 
The \textbf{low accuracy} classifier is essentially equal to \High, but trained on a training set with inversed labels. \Low's accuracy on the test set with non-inversed labels is 0.03.

\subsubsection*{Explanations}
For generating explanations, we focus on input features (single words) and influence on the prediction, as suggested in the minimum explanation setup by \cite{goodman16eu}. Following \cite{feng2018pathologies}, we highlight the most decisive words in the texts by colour. To convey just enough explanation, we highlight between $\frac{1}{3}$ and $\frac{1}{4}$ of the texts. The Tweets contain on average between 14 and 15 words, which results in $k=4$ highlighted words per Tweet. To generate \textbf{high-fidelity explanations} for \High and its inverse-label counterpart \Low, we use the L2X algorithm suggested by \cite{chen2018learning} on top of the CNN to select the most decisive features. For \Medium, we use the learned model coefficients. The \textbf{low-fidelity explanations} should not provide useful information about the underlying model, but should only be visually similar to the high-fidelity explanations. To generate such nonsensical explanations, we draw words uniformly at random from the texts.\newline
As the explanation is not tied to the classification result, the system can also show \textbf{no explanations} by not highlighting any word ($k=0$) but still show the classifier's prediction.

\subsubsection*{Subset Sampling}
It is not feasible to show the complete dataset to the participants during the user study, we therefore display a subset of 15 Tweets. To avoid affects from specific wording or topics in the subset, we generate subsets by drawing 15 Tweets at random from the test set. The subset is only kept if it is non-overlapping with previously drawn subsets, has a class balance similar to the test set, and if the classifiers' accuracies are equal to those on the test set. We then select the 10 subsets with the closest feature distribution compared to the training set, using the Kullback-Leibler Divergence (KLD) with Laplace smoothing (k=1). 

\subsubsection*{Explanation Evaluation}
For computer-generated explanations, it is possible that (1) the explanations constructed to have a high fidelity are meaningful to humans but are not faithful to the model, that (2) low-fidelity explanations nonetheless convey information about the classifier, and that (3) the explanations in the subsets show a different fidelity as those in the whole test set.\newline
To validate that the selected features are an actual representation of the classifier's reasoning, we reduce the texts of the test set to the $k$ selected features and subsequently let the respective classifier predict the label. If the explanations have a high fidelity to the underlying model, the reduced texts should lead to the same predictions as the original texts. We use the prediction for the original texts as ground truth for the reduced texts to calculate the label agreement. We repeat the evaluation for each subset to confirm that the fidelities of the explanations in the subsets do not differ from those of the complete test set.
% The IJCAI formatting rules specify that all tables and figures have to be floating at the top or bottom of the page, unless they are an integral part of the narrative (the latter might be justifiable for table 1)
\begin{table}[t]
	\centering
	\begin{tabular}{l|rr|rr|rr}
		& \HH & \HL & \MH & \ML & \LH & \LL \\ \midrule
		$k=1$  		&  0.97  &  0.58 & 1.00 & 0.64 & 0.97  &  0.59  \\
		\textbf{$k=4$}  &  \textbf{0.98}  &  \textbf{0.74} & \textbf{1.00} & \textbf{0.77} & \textbf{0.97}  &  \textbf{0.72}  \\
		$k=all$  	& 1.00	& 1.00 & 1.00	& 1.00	& 1.00	& 1.00  \\\midrule
		$\bar{x}_{subs}$	& 0.97 & 0.74 & 1.00 & 0.64 & 0.97 & 0.74 \\
		$s_{subs}$ 	& 0.03	& 0.13 & 0.00	& 0.12	& 0.03	& 0.10	\\\bottomrule
	\end{tabular}
	\caption{Label agreements evaluating the fidelity of explanations. Showing the accuracy of reduced texts when prediction of complete text is set as ground-truth for test set (top) and subsets (bottom, $k=4$). Class balance of 50:50 for both the test set and each subset.}
	\label{tab:exp2}
\end{table}
Table \ref{tab:exp2} shows that the high-fidelity explanations are enough to reproduce the original prediction of all three classifiers, even when reducing the texts to a single word. The low-fidelity explanations, on the contrary, cannot reliably reproduce the original predictions. We conclude that (1) the high-fidelity explanations indeed faithfully represent the underlying classifier, and that (2) a random selection of words is not explanatory for the classifiers.\newline
On average, the mean fidelities ($\bar{x}_{subs}$) of the subsets cor2respond to those on the complete test set. Only few subsets show differing fidelities, e.g. for \HL, two sets have a fidelity lower than the average (0.53), while one subset shows a higher fidelity level (0.93). The standard deviation of the subset fidelities ($s_{subs}$) are higher for randomly selected explanations than for high fidelity explanations, which was to be expected.

\subsubsection*{Graphical User Interface}
\begin{figure}[b]
	\includegraphics[width=\linewidth]{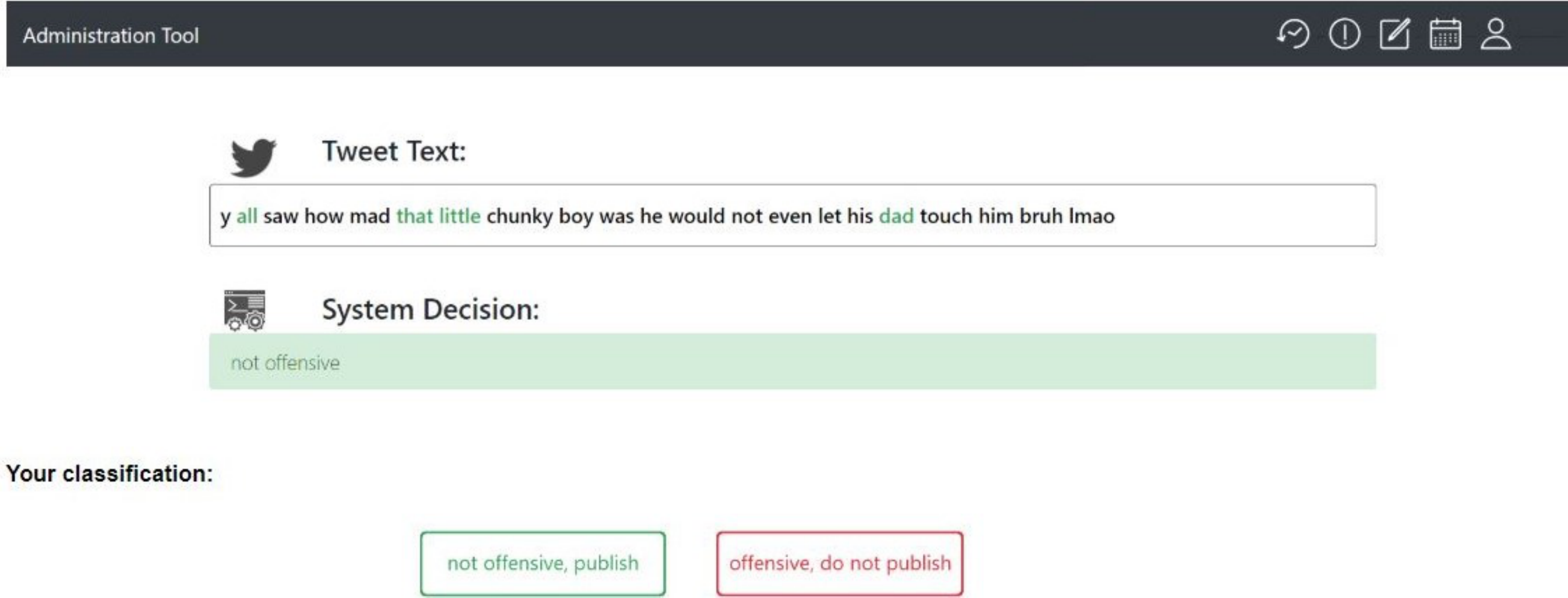}
	\caption{Screenshot of the graphical user interface in the survey}
	\label{fig:screenshot}
\end{figure}

For testing the effect of explanations of an automatic decision tool on users, we create an authentic and modern web interface with a minimalistic design, as to not distract the user from the main task. Figure \ref{fig:screenshot} shows the ``Administration Tool", a software tool to support a social media administrator in detecting offensive content. 
%The tool is a means to display the texts, the explanations (for the high-fidelity and low-fidelity conditions), as well as the classifier's recommendation. Furthermore, the user should believe that the system is capable of providing intelligent functionality. We therefore use the front-end web framework \textit{Bootstrap}\footnote{https://getbootstrap.com accessed on 24.02.2019} to generate a modern and responsive web interface with a minimalistic design, as to not distract the user from the main task.

%--------------------------------------------------------------------------------------
\section{Results and Discussion}
\label{sec:res_disc}

\begin{figure}[t]
	\includegraphics[width=\linewidth]{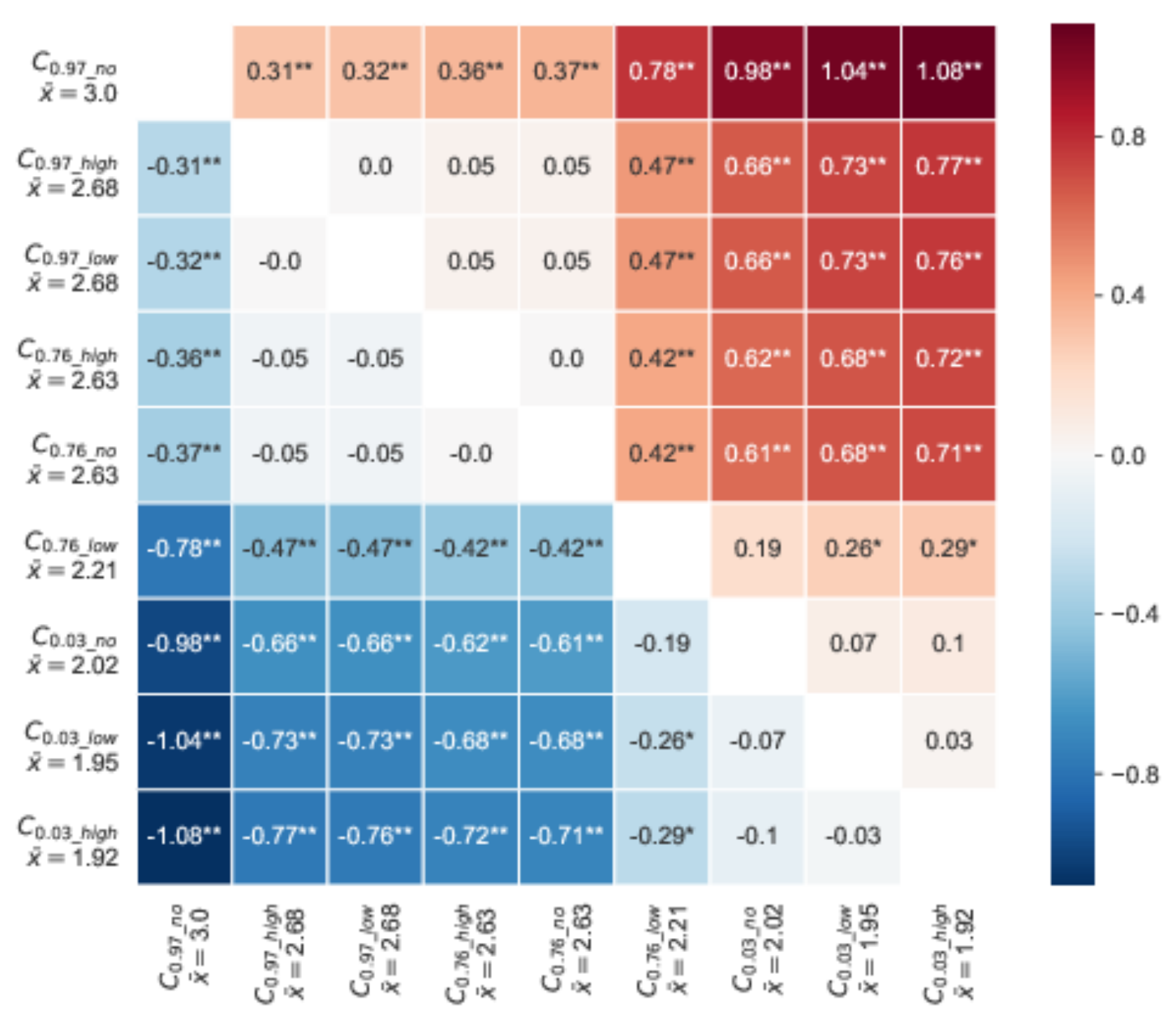}
	\caption{Comparison of self-reported trust scores ordered by mean ($\bar{x}$), value reporting difference of means ($\bar{x}_{row} - \bar{x}_{column}$ ), asterisk reporting significance (* significant at $\alpha=\frac{0.05}{9}$, ** significant at $\alpha=\frac{0.01}{9}$)}
	\label{fig:matrix_trust}
\end{figure}

\begin{figure}[t]
	\includegraphics[width=\linewidth]{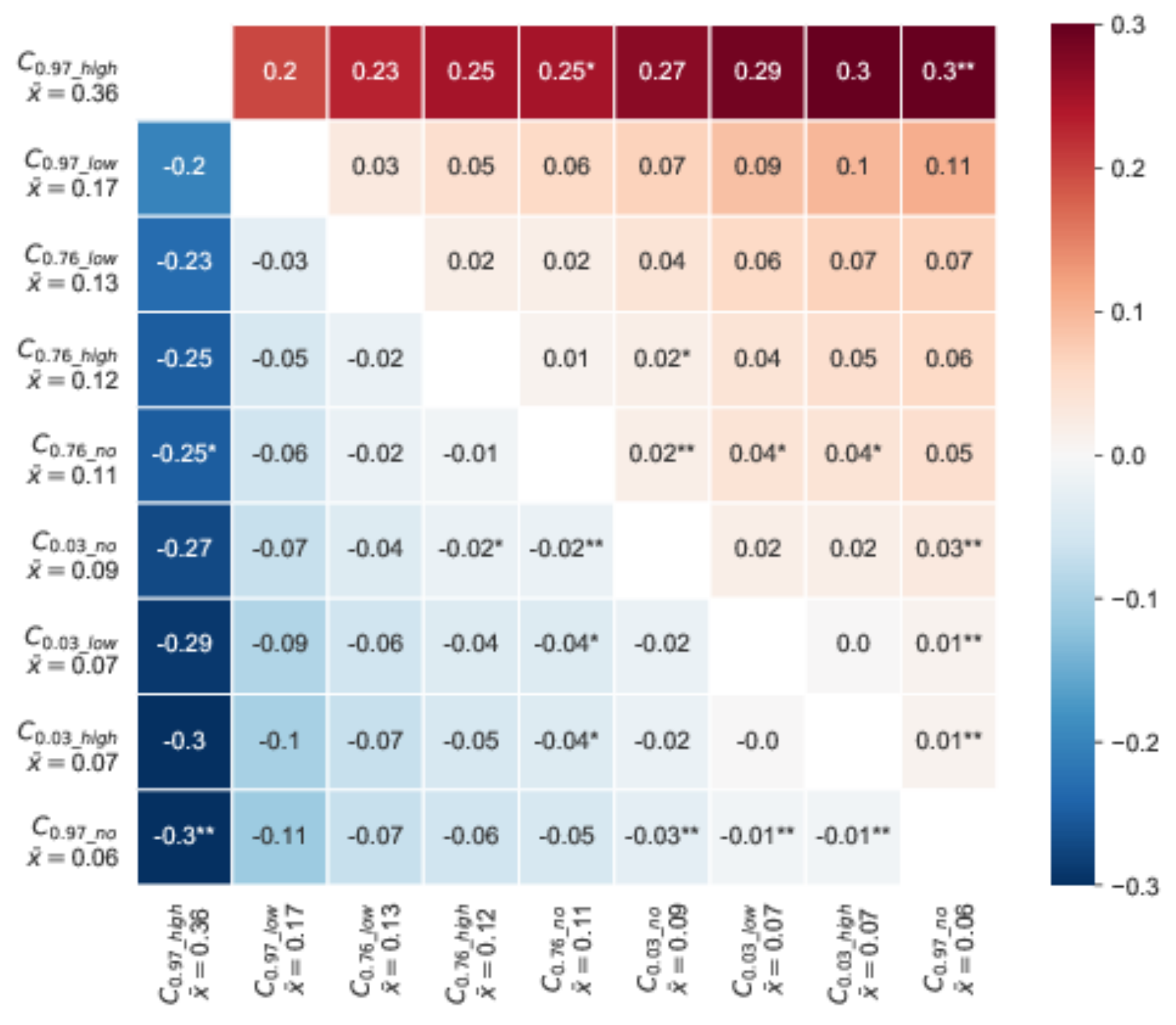}
	\caption{Comparison of observed trust scores (relative changes towards classifier away from truth) ordered by mean, value reporting difference of means ($\bar{x}_{row} - \bar{x}_{column}$ ), asterisk reporting significance (* significant at $\alpha=\frac{0.05}{9}$, ** significant at $\alpha=\frac{0.01}{9}$)}
	\label{fig:matrix_rel_changes}
\end{figure}

Table \ref{tab:results_table} reports the mean scores and their standard deviations for \emph{self-reported trust, perceived understanding}, and \emph{perceived predictability} (an individual item from the self-reported trust questionnaire). Figures \ref{fig:matrix_trust}-\ref{fig:matrix_rel_changes} show the differences in means between all conditions and the significance of the sample comparisons (denoted with an asterisk) for self-reported trust and \emph{observed trust} trust. All significance scores use Bonferroni correction to account for the multiple comparisons bias. Table \ref{tab:conf_matrix} presents the results of observed trust in changes towards or away from the truth and the classifier's prediction, respectively.
% The IJCAI formatting rules specify that all tables and figures have to be floating at the top or bottom of the page, unless they are an integral part of the narrative (the latter might be justifiable for table 1)
%\begin{figure}[t]
%	\includegraphics[width=\linewidth]{img/matrix_understanding.eps}
%	\caption{Comparison of perceived understanding scores ordered by mean, value reporting difference of means ($\bar{x}_{row} - \bar{x}_{column}$ ), asterisk reporting significance (* significant at $\alpha=\frac{0.05}{9}$, ** significant at $\alpha=\frac{0.01}{9}$)}
%	\label{fig:matrix_understanding}
%\end{figure}
\begin{table}[b]
\centering
	\begin{tabular}{+l^c^cc}
	\toprule\tabhead
		 & \textbf{self-rep.} & \multicolumn{2}{c}{\textbf{perceived}} \\ 
		 \tabhead
		Condition & \textbf{trust} & \textbf{underst.} & \textbf{predict.} \\ \midrule
		\vspace{0.05cm}$C_{0.97\_high}$& 2.7  $\pm$  0.4 & 3.9  $\pm$  0.9 & 3.0  $\pm$  0.8  \\
		\vspace{0.05cm}$C_{0.97\_low}$& 2.7  $\pm$  0.5 & 3.7  $\pm$  0.9 & 2.9  $\pm$  0.8  \\
		\vspace{0.05cm}$C_{0.97\_no}$& 3.0  $\pm$  0.5 & 4.1  $\pm$  0.7 & 3.2  $\pm$  0.7  \\
		
		\vspace{0.05cm}$C_{0.76\_high}$& 2.6  $\pm$  0.5 & 3.8  $\pm$  0.8 & 2.9  $\pm$  0.6  \\
		\vspace{0.05cm}$C_{0.76\_low}$& 2.2  $\pm$  0.5 & 2.9  $\pm$  1.0 & 2.3  $\pm$  0.8  \\
		\vspace{0.05cm}$C_{0.76\_no}$& 2.6  $\pm$  0.5 & 3.7  $\pm$  0.7 & 2.7  $\pm$  0.8  \\
		
		\vspace{0.05cm}$C_{0.03\_high}$& 1.9  $\pm$  0.4 & 2.5  $\pm$  1.2 & 2.0  $\pm$  0.6  \\
		\vspace{0.05cm}$C_{0.03\_low}$& 2.0  $\pm$  0.4 & 2.5  $\pm$  1.1 & 1.8  $\pm$  0.7  \\
		\vspace{0.05cm}$C_{0.03\_no}$& 2.0  $\pm$  0.5 & 2.9  $\pm$  1.2 & 2.1  $\pm$  0.8  \\ \bottomrule
	\end{tabular}
	\caption{Means and standard deviations for self-reported trust, perceived understanding, and predictability scores.}
	\label{tab:results_table}
\end{table}

%\begin{table}[H]
%	\centering
%	\begin{tabular}{ll|rr}
%		\multicolumn{2}{l|}{} & \multicolumn{2}{c}{\textbf{Truth}}\\
%		\multicolumn{2}{l|}{} & Towards & Away \\ \midrule
%		
%		\textbf{High-} & Towards & 32 & 4 \\
%		\textbf{highf} & Away & 0 & 6 \\ \midrule
%
%		\textbf{High-} & Towards & 22 & 2 \\
%		\textbf{lowf} & Away & 0 & 11 \\ \midrule
%		
%		\textbf{High-} & Towards & 31 & 1 \\
%		\textbf{no} & Away & 0 & 8 \\ \midrule
%		
%		\textbf{Medium-} & Towards & 22 & 10 \\
%		\textbf{highf} & Away & 3 & 2 \\ \midrule
%		
%		\textbf{Medium-} & Towards & 17 & 9 \\
%		\textbf{lowf} & Away & 3 & 12 \\ \midrule
%		
%		\textbf{Medium-} & Towards & 20 & 7 \\
%		\textbf{no} & Away & 1 & 2 \\ \midrule
%		
%		\textbf{Low-} & Towards & 1 & 27 \\
%		\textbf{highf} & Away & 7 & 1 \\ \midrule
%		
%		\textbf{Low-} & Towards & 1 & 23 \\
%		\textbf{lowf} & Away & 10 & 0 \\ \midrule
%		
%		\textbf{Low-} & Towards & 2 & 31 \\
%		\textbf{no} & Away & 4 & 0 \\ \midrule
%	\end{tabular}
%	\caption{Influence of systems on user labelling behaviour: absolute changes when confronted with system prediction, per classifier}
%\end{table}

\begin{table}[t]
	\centering
	\begin{tabular}{+l^l^l^r^r}
	\toprule
		&\multicolumn{2}{l}{} & \multicolumn{2}{c}{\textbf{Truth}}\\
		&\multicolumn{2}{l}{} & Towards & Away \\ \otoprule
 \multirow{20}{*}{\rotatebox[origin=c]{90}{\textbf{Classifier Prediction}}}	
 
		\vspace{-0.06cm}&\multirow{2}{*}{\HH} & Towards & 0.32 & 0.36 \\
		\vspace{-0.06cm}& & Away & 0.00 & 0.02 \\ \cmidrule(r){2-5}
					
		\vspace{-0.06cm}&\multirow{2}{*}{\HL} & Towards & 0.19 & 0.17 \\
		\vspace{-0.06cm}& & Away & 0.00 & 0.03 \\ \cmidrule(r){2-5}
		
		\vspace{-0.06cm}&\multirow{2}{*}{\HN} & Towards & 0.26 & 0.06 \\
		\vspace{-0.06cm}& & Away & 0.00 & 0.02 \\ \cmidrule(r){2-5}
		
		\vspace{-0.06cm}& \multirow{2}{*}{\MH} & Towards & 0.27 & 0.12 \\
		\vspace{-0.06cm}& & Away & 0.09 & 0.01 \\ \cmidrule(r){2-5}
		
		\vspace{-0.06cm}&\multirow{2}{*}{\ML} & Towards & 0.19 & 0.13 \\
		\vspace{-0.06cm}& & Away & 0.19 & 0.05 \\ \cmidrule(r){2-5}
		
		\vspace{-0.06cm}&\multirow{2}{*}{\MN} & Towards & 0.33 & 0.11 \\
		\vspace{-0.06cm}& & Away & 0.05 & 0.01 \\ \cmidrule(r){2-5}
		
		\vspace{-0.06cm}&\multirow{2}{*}{\LH} & Towards & 0.17 & 0.07 \\
		\vspace{-0.06cm}& & Away & 0.11 & 0.07 \\ \cmidrule(r){2-5}
		
		\vspace{-0.06cm}&\multirow{2}{*}{\LL} & Towards & 0.25 & 0.07 \\
		\vspace{-0.06cm}& & Away & 0.10 & 0.00 \\ \cmidrule(r){2-5}
		
		\vspace{-0.06cm}&\multirow{2}{*}{\LN} & Towards & 0.33 & 0.09 \\
		\vspace{-0.06cm}& & Away & 0.04 & 0.00 \\ \bottomrule
	\end{tabular}
	\caption{Influence of systems on user labelling behaviour: relative changing frequencies when confronted with system prediction, per classifier, normalised over opportunities.}
	\label{tab:conf_matrix}
\end{table}

\subsection{Model Accuracy}
Our results suggest that model accuracy has a stronger influence on user trust than explanation fidelity. Figure \ref{fig:matrix_trust} shows that when ordered by trust score, no system of \Low~is ranked higher than any system of \Medium~or \High, and no system of \Medium~is ranked higher than any system of \High. These findings are in line with the research of \cite{yu2017user}. It also aligns with the ``expectation mismatch" described in \cite{glass2008toward}: A classifier with high accuracy leads to fewer mismatches with the user's expectations, which in turn does not decrease the trust. We observe the same trend for user ratings of predictability: low accuracy systems are rated to be less predictable than high accuracy systems. Both classifiers, however, are objectively equally predictable, since they behave exactly the same (\Low always returns the opposite label from \High; their results hence have equal entropy). This suggests that user's perception is heavily influenced by accuracy levels.

\subsection{Explanations}
In our experiment, the presence of an explanation did not have a positive effect on self-reported trust in any of the conditions (figure \ref{fig:matrix_trust}). Adding an explanation to the system decreased the trust in the case of \High~and did not influence trust in \Low. For \Medium, the type of explanation was crucial for its influence -- a high-fidelity explanation did not decrease trust levels significantly, while a low-fidelity explanation did.\newline
For \High, \HN~shows better results than \HH~and \HL. With \HN, there is no ``expectation mismatch" as no explanation is given and accuracy is high. The explanations of \HH, however, are built on statistical rather than causal relations, while \HL's explanations are random. As humans make sense of new observation by using previously learned knowledge, i.e. assuming human-like reasoning strategies even for an algorithm, seeing any of those two explanations leads to a deceptive experience. Contrarily, the observed trust measure (figure \ref{fig:matrix_rel_changes}) shows a significantly lower score for \HN~than for \HH, meaning that participants show a higher willingness to accept the predictions of \HH~than for \HN. \cite{langer1978mindlessness} noticed a difference between a ``mindless" (non-attentive) and a ``mindful" (attentive) state, which could be the explanation for the difference between a self-reported (attentively) and an observed (non-attentively) measure. Users do not report different trust levels for \HH~and \HL, but they more often follow \HH's recommendation than \HL's (table \ref{tab:conf_matrix}) -- their behaviour is hence influenced by the level of truthfulness.\newline
Unlike \High, \Medium~shows an equal self-reported trust score for \MH~and \MN~and a significantly lower score for \ML. Making three to four mistakes on each subset, it is imaginable that users are more conscious about the classifier's behaviour than they are with \High~due to the higher error rate. Having at least an indication of the reasons for misclassifications (\MH) could in turn increase the trust. For \ML, the ``expectation mismatch" is twofold, bringing together misclassifications and nonsensical explanations. Looking at the observed trust, \MH~has the highest trust rate, while \ML~has the highest rate of mistrust.\newline
\Low~did not show evidence of diverting self-reported trust scores for any of the three explanation types. The same homogeneity is found in the observed trust scores, for both trust and mistrust. This suggests that users are not fooled by a bad classifier and do not trust it, no matter the explanation given.

\subsection{Objective Trust Measure}
Self-reporting requires users to have the ability to reflect on and process their relationship with the system. Using an objective measure for trust avoids the necessity of this ability. In our results, we see that the observed (hence potentially unconscious) trust scores do not always align with the self-reported trust scores. Although users of \HN~have the highest self-reported trust score, they are not as easily ``lured" towards a wrong classification as users of \HH. If this is due to an actual gap between actions and reflections of users, the observation measure could be interesting for xAI practitioners as it shows how users actually interact with a system. However, as our results of the observed trust measure are ambiguous and have high variance, the measure should be validated in future research.

\subsection{Limitations}
In this study, we make use of minimum explanations which show the relation between the input and output but do not deliver information about the inner structure of a classifier. The influence of the task (difficulty) and explanation visualisation (detailedness) should be further investigated. It should also be tested in future research whether users accept only high-fidelity explanations, or likewise accept explanations that look meaningful to a human but are not faithful to the underlying machine learning algorithm. The study results are furthermore limited by the cultural background of the participants (mainly Caucasians). The results therefore cannot be generalised across cultural backgrounds and the connected general attitude towards technology.

%--------------------------------------------------------------------------------------
\section{Conclusion}

This paper presents empirical evidence for the impact of model accuracy and explanation fidelity on user trust. We generated minimal explanations with high and low fidelity for three systems with different performance levels. We then validated the explanations' fidelity level and tested differences in nine conditions (3 model accuracy levels x 3 explanation fidelity levels) in a user study.\newline
Our findings show that explanations affect user trust in a variety of ways, depending on the overall accuracy of the system, the fidelity of the explanation, and the user's level of consciousness. Participants showed the most trust in systems without explanations, i.e. minimum explanations can potentially harm, but not improve user trust. We argue that the act of reconciling conflicting information of the mental model and the given explanations counts as a deceptive experience and therefore affects the user's trust negatively. If an explanation is added to a system (e.g. for increasing user's understanding of the system), its fidelity is crucial for user trust. We saw that for systems with a medium accuracy (\Medium), a high-fidelity explanation does not harm user trust, while a low-fidelity explanation decreases trust. Overall, the model's accuracy levels showed the most impact on trust levels. We furthermore found that users' awareness level influences their perception of trust. The results found from self-reported trust measures show a different picture than when objectively observing trust via the participant's actions.\newline
Further research with more rich explanations and a detailed investigation of trust factors is needed to examine potential positive effects of explanations on user trust. The development of trust over time should also be researched in the future, to give practical directions to xAI practitioners implementing explanations in productive systems.

%--------------------------------------------------------------------------------------

%\section*{Acknowledgments}

%--------------------------------------------------------------------------------------
%% The file named.bst is a bibliography style file for BibTeX 0.99c
\clearpage
\bibliographystyle{named}
\bibliography{ijcai19}

\end{document}